\documentclass[aps,longbibliography,showpacs,twocolumn,superscriptaddress,amsmath,amssymb,verbatim]{revtex4-1}

\usepackage[version=4]{mhchem}

\usepackage[thinlines]{easytable}
\usepackage{graphicx}
\usepackage{lmodern}
\usepackage{epstopdf}
\usepackage{dcolumn}
\usepackage{bm}
\usepackage{subfigure}
\usepackage{makecell}
\usepackage{amsmath} 
\usepackage[pdftex,colorlinks=true,citecolor=blue,linkcolor=blue,urlcolor=blue,bookmarks=true]{hyperref}
\usepackage{booktabs}
\usepackage{multirow} 
\usepackage{array} 

\usepackage{color}
\usepackage{textcomp}
\definecolor{red}{rgb}{1,0,0}

\definecolor{blue}{rgb}{0,0,1}

\definecolor{green}{rgb}{0,1,0}

\begin{document}
	\preprint{APS}

\title{\textbf{ Magnetism and spin dynamics of Na\textsubscript{5}Yb(MoO\textsubscript{4})\textsubscript{4}: A weakly interacting rare-earth stretched diamond lattice}} 
\author{N. Rajeesh Kumar}
\email{rajeeshphysik@gmail.com}
\affiliation {\it Center for Condensed Matter Sciences, National Taiwan University, Taipei 10617, Taiwan}
\affiliation {\it \textit{Institute of Physics, Academia Sinica, Taipei 10617, Taiwan} }
\author{J. Khatua}
\affiliation {\it Department of Physics, Sungkyunkwan University, Suwon 16419, Republic of Korea}
\author{Changhyun Koo}
\affiliation {\it Department of Physics, Sungkyunkwan University, Suwon 16419, Republic of Korea}
\author{Izumi Umegaki}
\affiliation {\it Muon Science Laboratory, IMSS, KEK, Tokai, Ibaraki 319-1112, Japan}
\author{C. -E. Yin }
\affiliation {\it Department of Physics, National Tsing Hua University, Hsinchu 30013, Taiwan}
\author{C. -W. Wang }
\affiliation {\it National Synchrotron Radiation Research Center, Hsinchu 300092, Taiwan}
\author{A. M. Strydom }
\affiliation {\it \textit{Highly Correlated Matter Research Group, Department of Physics, University of Johannesburg, Auckland Park 2006, South Africa} }
\author{H. -T. Jeng }
\affiliation {\it Department of Physics, National Tsing Hua University, Hsinchu 30013, Taiwan}
\author{Kwang-Yong Choi}
\affiliation {\it Department of Physics, Sungkyunkwan University, Suwon 16419, Republic of Korea}
\author{R. Sankar}
\email{sankarraman@gate.sinica.edu.tw}
\affiliation {\it \textit{Institute of Physics, Academia Sinica, Taipei 10617, Taiwan} }
\affiliation{\it Taiwan Consortium of Emergent Crystalline Materials, National Science and Technology Council, Taipei 10622, Taiwan}
\author{W. -T. Chen}
\email{weitinchen@ntu.edu.tw}
\affiliation {\it Center for Condensed Matter Sciences, National Taiwan University, Taipei 10617, Taiwan}
\affiliation{\it Taiwan Consortium of Emergent Crystalline Materials, National Science and Technology Council, Taipei 10622, Taiwan}
\affiliation {\it Center of Atomic Initiative for New Materials, National Taiwan University, Taipei 10617, Taiwan}

\date{\today}

\begin{abstract}
	We report a comprehensive investigation of the structural and magnetic properties of Na$_5$Yb(MoO$_4$)$_4$, a member of the stretched diamond magnetic lattice family. Neutron powder diffraction at 3.3~K confirms that the compound crystallizes in the tetragonal \textit{I4$_1$/a} space group, with a large interatomic separation of 6.33~\AA{} between magnetic Yb ions forming a three-dimensional stretched diamond framework. Magnetic susceptibility and specific heat measurements reveal no evidence of long-range magnetic order down to 60~mK. The low-temperature magnetic behavior is governed by an effective $J_{\mathrm{eff}} = 1/2$ Kramers doublet ground state, well separated from excited crystal-field levels, arising from the distorted dodecahedral oxygen coordination of Yb$^{3+}$. Density functional theory calculations within the DFT+$U$ framework indicate that exchange interactions between Yb ions are negligibly small, consistent with the long O--Mo--O super-superexchange pathways. The temperature dependence of the specific heat exhibits signatures of gapped spin excitations, most likely originating from long-range dipolar correlations and further shaped by weak exchange interactions together with the strong single-ion anisotropy of the Yb moments. Muon spin relaxation measurements reveal persistent low-energy spin dynamics, indicating that dipolar correlations remain dynamic and are insufficient to stabilize static magnetic order down to 50~mK. These results identify Na$_5$Yb(MoO$_4$)$_4$ as a rare example of a dipolar quantum paramagnet in which single-ion physics and long-range dipolar interactions dominate, while exchange interactions are suppressed to the millikelvin energy scale.
\end{abstract}
\maketitle
\section{Introduction}

In condensed matter physics, quantum-disordered ground states are characterized by the absence of long-range order down to absolute zero temperature, arising from strong quantum fluctuations that persist even at \(T=0\) K~\cite{Balents2010,Savary_2016}. A wide variety of transition-metal- and rare-earth-based compounds have been extensively investigated to realize such ground states, providing a fertile platform to explore novel quantum phenomena. Among them, the magnetism of rare-earth materials is particularly intriguing as it originates from their localized $4f$ electrons, where the interplay of strong spin--orbit coupling and crystal electric field (CEF) effects  gives rise to pronounced magnetic anisotropy and relatively weak exchange interactions between magnetic ions~\cite{doi:10.1146/annurev-conmatphys-020911-125138}. These unique characteristics, when combined with suitable lattice geometries, can facilitate geometric frustration and give rise to a wide variety of unconventional magnetic ground states, including non-collinear magnetic order~\cite{PhysRevB.92.144423}, Bose--Einstein condensation~\cite{PhysRevLett.123.027201,CFeng2023}, spin ice~\cite{science1064761,Castelnovo2008}, quantum spin liquids~\cite{Arh2022,Scheie2024}, quantum dipolar magnets~\cite{PhysRevB.104.L220403,PhysRevB.104.024427} and dimerized phases~\cite{Liu2025}.\par
A diamond magnetic lattice is a bipartite network and, in its ideal form, does not exhibit frustration, stabilizing a conventional N\'eel antiferromagnetic or an incommensurate spiral ground state \cite{oitmaa2019frustrated,PhysRevB.96.064413}. However, distorted or ``stretched'' diamond lattices have recently gained attention as potential hosts of magnetic frustration~\cite{PhysRevMaterials.6.044410}. Although the stretching preserves the bipartite nature of the lattice, frustration can emerge from competition between nearest-neighbor ($J_1$) and next-nearest-neighbor ($J_2$) exchange interactions. Several material systems, including rare-earth tantalates RETaO$_4$ \cite{PhysRevMaterials.6.044410} and niobates RENbO$_4$ \cite{PhysRevB.111.014411}, as well as alkali-based oxides such as NaCeO$_2$ and LiYbO$_2$ \cite{PhysRevLett.130.166703,PhysRevB.103.024430}, serve as prominent examples where magnetic frustration originates from competing $J_1$--$J_2$ interactions on a stretched diamond lattice. As a consequence of this frustration, exotic quantum phases such as quantum spin liquids~\cite{Paddison2017}, spiral spin liquids~\cite{Bergman2007}, and topological paramagnets have been realized in these systems \cite{PhysRevLett.130.166703,PhysRevB.108.134424,PhysRevB.111.014411}.

In this context, we investigate the magnetic ground state of the rare-earth molybdate compound Na$_5$Yb(MoO$_4$)$_4$, which crystallizes in a stretched diamond magnetic lattice. This compound can be viewed as a complex derivative of the conventional scheelite-type ABO$_4$ structure, crystallizing in the tetragonal space group \textit{I4$_1$/a} \cite{DHANYA20186699,chen2025}. The magnetic lattice in Na$_5$Yb(MoO$_4$)$_4$ is highly unusual, featuring a remarkably large nearest-neighbor Yb--Yb separation of approximately 6.3~\AA. The magnetic interactions are mediated through extended super-superexchange pathways, in contrast to previously studied stretched-diamond systems where the $J_1$ distance typically lies in the range of 3--5~\AA\ and is predominantly governed by superexchange interactions. Furthermore, the next-nearest-neighbor ($J_2$) Yb--Yb distance exceeds 9~\AA, significantly weakening the $J_2$ exchange interactions. As a result, magnetic frustration arising from competing $J_1$–$J_2$ exchange is drastically suppressed in Na$_5$Yb(MoO$_4$)$_4$, distinguishing it from other frustrated diamond-lattice systems.\par
We employ neutron and synchrotron X-ray diffraction to elucidate the structural details of Na$_5$Yb(MoO$_4$)$_4$. The magnetic properties and ground state are investigated using bulk magnetic susceptibility measurements, specific heat studies, and muon spin relaxation ($\mu$SR) experiments. In addition, density functional theory (DFT) calculations within the DFT+$U$ framework are used to provide theoretical support for the experimental findings. Our results establish Na$_5$Yb(MoO$_4$)$_4$ as a quantum paramagnet in which the exchange interactions are negligible and dipolar interactions dominate the magnetic energy scale. Similar dipolar-dominated quantum disordered ground states are observed in  several rare-earth-based geomterically frustrated compounds such as Yb(BaBO$_3$)$_3$~\cite{PhysRevB.106.014409}, Gd$_2$Sn$_2$O$_7$~\cite{PhysRevLett.99.097201}, Yb$_3$Ga$_5$O\textsubscript{12}~\cite{PhysRevB.104.024427} and NaGdS$_2$~\cite{PhysRevB.111.144404}. However, in our case,  quantum disorder arises on an unfrustrated magnetic lattice with ultraweak exchange coupling, dominated by Yb single ion anisotropy and long-range dipolar interactions.
\section{Research methods}
Polycrystalline samples of the title compound $\mathrm{Na_5Yb(MoO_4)_4}$ were synthesized by a conventional solid-state method. High-purity chemicals, including Yb$_{2}$O$_{3}$, MoO$_{3}$, and Na$_{2}$CO$_{3}$ with purity $>$ 99.9 \text{\%}, were purchased from Sigma-Aldrich. Yb$_{2}$O$_{3}$ (Y$_{2}$O$_{3}$) was preheated at 1073 K for 24 h, while Na$_{2}$CO$_{3}$ was dried at 393 K prior to use. Stoichiometric mixtures of the starting materials were thoroughly ground using an agate mortar and subjected to sequential heat treatments at 873 K and 923 K for 24 h each, with intermediate regrinding to ensure homogeneity. Reaction temperatures above 923 K resulted in decomposition of the product.\\\begin{figure}[b]
	\centering
	\includegraphics[width=0.48\textwidth]{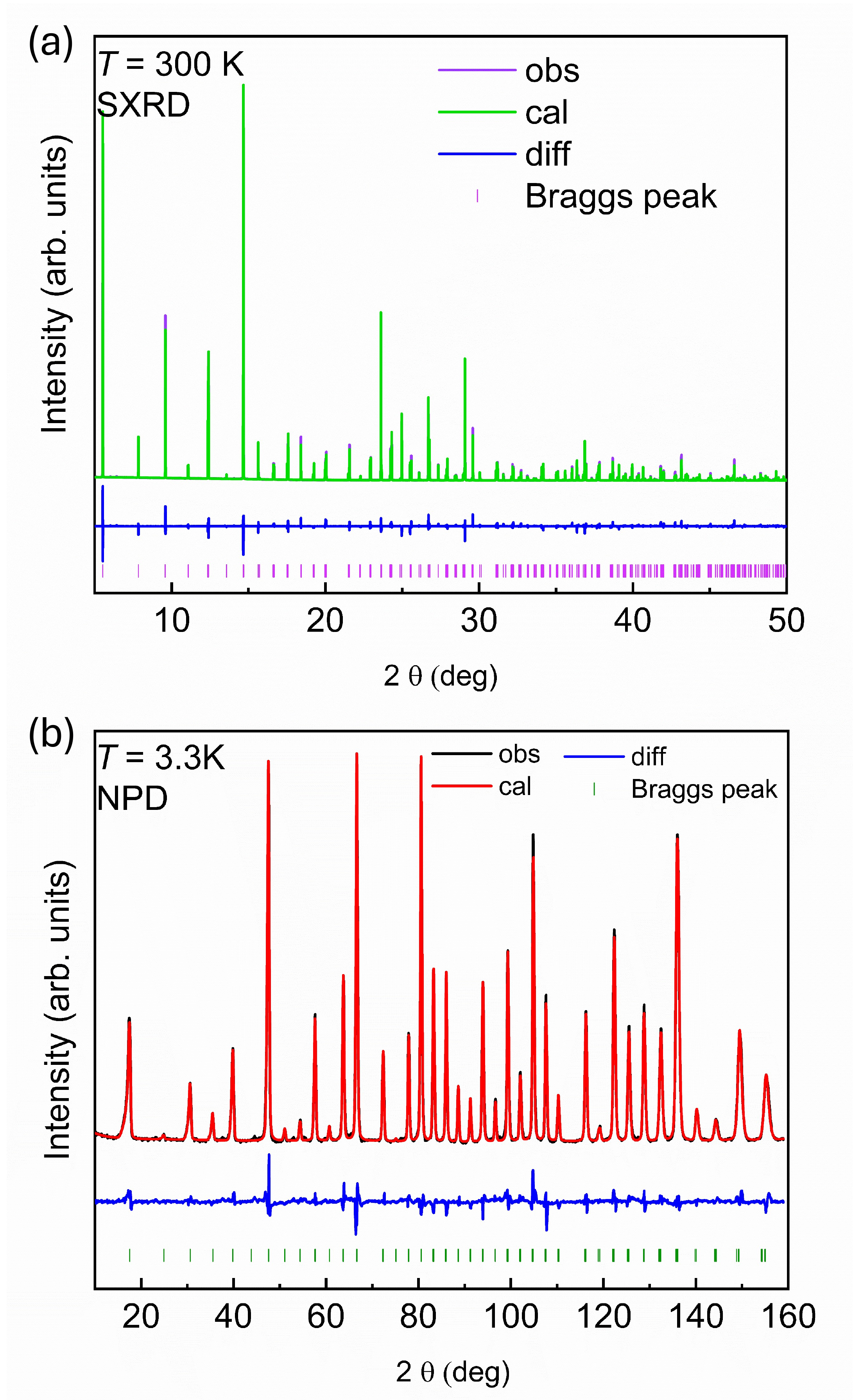}
	\caption{(a) Synchrotron X-ray diffraction pattern of \ce{Na5Yb(MoO4)4} collected at 300~K.
		(b) Neutron diffraction pattern of \ce{Na5Yb(MoO4)4} acquired at 3.3~K.
		The legends in both panels indicate the corresponding meaning of the color codes.}
	\label{NPD}
\end{figure}
\begin{figure*}[t]
	\centering
	\includegraphics[width=\textwidth]{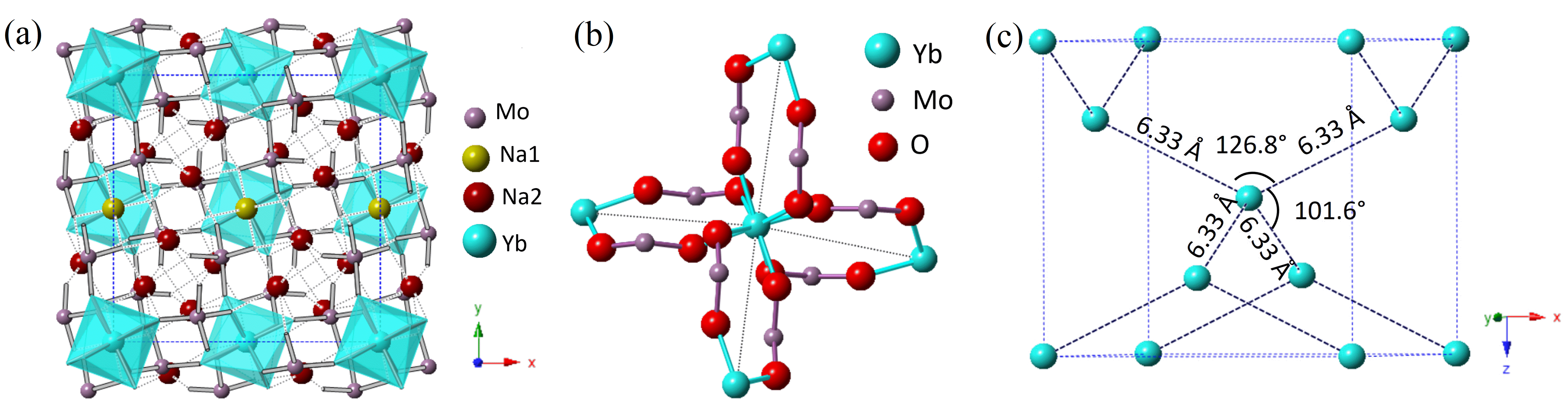}
	\caption{(a) Crystal structure of \ce{Na5Yb(MoO4)4} viewed along the $y$-axis.
		(b) Nearest-neighbor Yb atoms connected through O--Mo--O super-superexchange pathways.
		(c) Yb atoms forming a stretched diamond lattice along z-axis. The tetragonal unit cell is indicated by dotted blue lines.}
	\label{ST}
\end{figure*}
 High-resolution synchrotron X-ray diffraction (SXRD) measurements were conducted at the TPS 19A beamline of the National Synchrotron Radiation Research Center, Taiwan, using a 16 keV beam and a MYTHEN detector at 300 K. Neutron powder diffraction (NPD) data were collected on the WOMBAT high-intensity powder diffractometer at the Australian Nuclear Science and Technology Organisation (ANSTO) at $T = 3.3$ K. Rietveld refinement of the diffraction data was carried out using the Bruker TOPAS software package.\\ Magnetic measurements were performed in the temperature range $1.8 \leq T \leq 350$ K and in magnetic fields up to 7 T using a Magnetic Property Measurement System (MPMS3, Quantum Design, USA). Specific-heat measurements ($C_{\mathrm{p}}$) were carried out by the standard relaxation method using a Physical Property Measurement System (PPMS, Quantum Design, USA) in the range $1.8 \leq T \leq 200$ K and in fields up to 9 T. Sub-kelvin \textit{ac} magnetic susceptibility measurements were further performed down to 60 mK using a dilution refrigerator insert with the ac-susceptibility option in a Dynacool cryostat (Quantum Design, USA). Likewise, millikelvin specific-heat data were collected in the range $60~\mathrm{mK} \leq T \leq 4~\mathrm{K}$ using the heat-capacity option of a dilution refrigerator insert on the same Dynacool platform.
\\
The CEF analysis was performed using the pycrystalfield
python  package based on the point-charge model~\cite{Scheie:in5044,NewmanNg2000}.\\
Zero-field and longitudinal-field muon spin relaxation measurements were performed at the D1 beamline of the Materials and Life Science Experimental Facility (MLF), J-PARC, using a spin-polarized pulsed surface-muon ($\mu^{+}$) beam. A dilution refrigerator and a standard $^4$He flow cryostat were used to acquire data over a wide temperature range of $50~\mathrm{mK} \leq T \leq 98~\mathrm{K}$. The $\mu$SR spectra were analyzed using the MUSRFIT software package~\cite{SUTER201269}.
\\
First-principles calculations were performed using the Vienna \emph{Ab initio} Simulation Package (VASP)~\cite{PhysRevB.47.558,PhysRevB.49.14251,PhysRevB.54.11169,KRESSE199615} within the framework of DFT~\cite{Kohn1996}. The projector augmented-wave (PAW) method was employed together with the Perdew--Burke--Ernzerhof (PBE) exchange--correlation functional~\cite{PhysRevLett.77.3865}. A Hubbard $U$ correction of $U = 6.0$ eV was applied to the Yb $4f$ orbitals~\cite{Rk2015}.\\
A plane-wave cutoff energy of $650$~eV and a $\Gamma$-centered
Monkhorst--Pack~\cite{PhysRevB.13.5188} \emph{k}-point mesh of $2 \times 2 \times 2$ were used
to sample the Brillouin zone of the large unit cells with lattice constants
exceeding $11$~\AA. In addition, the \text{TB2J} code was employed to calculate
the magnetic exchange parameters within the dynamically stable approximation~\cite{HE2021107938}.\\
\begin{figure*}[t]
	\centering
	\includegraphics[width=\textwidth]{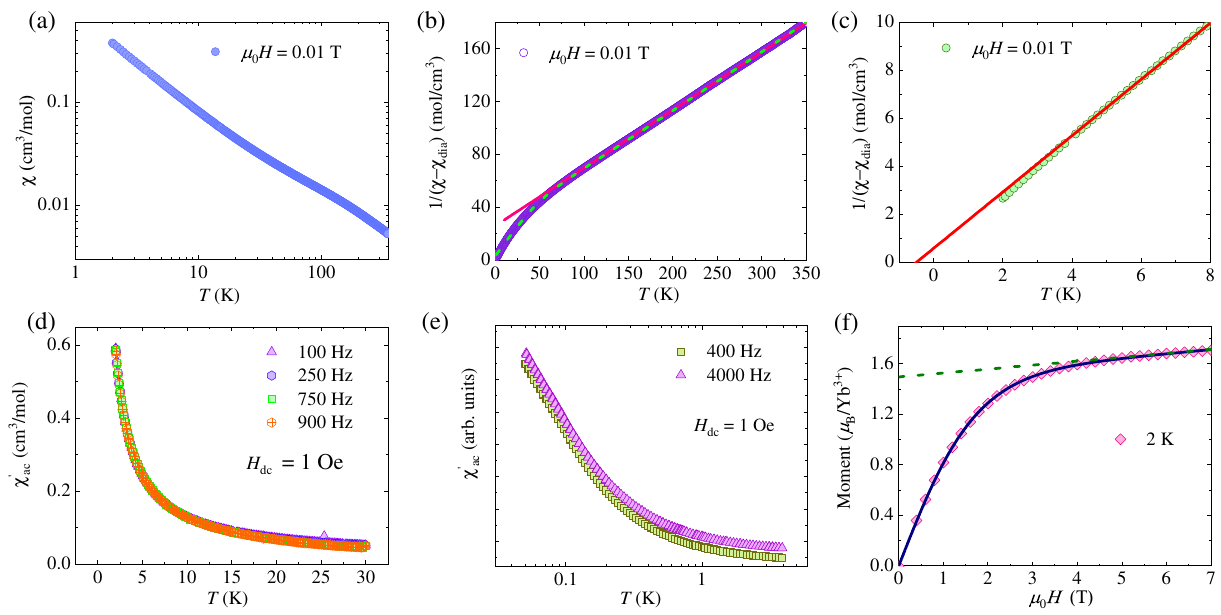}
	\caption{(a)Temperature dependence of magnetic susceptibility of $\mathrm{Na_5Yb(MoO_4)_4}$ acquired between 1.8 K and 300 K at $\mu_{0}H$ = 0.01 T. (b) Temperature dependence of  inverse magnetic susceptibility after subtracting the diamagnetic contribution. The solid pink line represents the Curie–Weiss (CW) fit in the high-temperature region, while the dashed blue line corresponds to the fit using Eq.~\ref{gapcef}. (c) Zoomed view of the inverse magnetic susceptibility in the low-temperature region with a CW fit (red line).   (d) \textit{ac} magnetic susceptibility in the temperature range \(1.8~\mathrm{K} \leq T \leq 30~\mathrm{K}\)  at several frequencies. (e)  Temperature dependence of \textit{ac} magnetic susceptibility in the temperature range $60$ mK $\leq$ $T$ $\leq$ 4 K at two different frequencies. (d)  Isothermal magnetization at \(T = 2\) K, where the solid dark blue line represents the combined fit using a Brillouin function and a linear Van Vleck term, while the dashed green line indicates the linear Van Vleck contribution.}
	\label{mag}
\end{figure*}
\section{Results} 
\subsection{Crystal structure}
The phase purity and structural stability of the synthesized polycrystalline sample were confirmed using room-temperature SXRD and low-temperature NPD measurements.  The Rietveld refinement profile of the SXRD data at room temperature is shown in Fig.~\ref{NPD}(a). The diffraction experiments do not reveal any detectable secondary phase. However, the specific-heat data suggest the presence of a small amount of unreacted impurity phase of $\mathrm{Yb_2O_3}$~(see Sec.~\ref{SP}). Figure~\ref{NPD}(b) represents the NPD pattern of $\mathrm{Na_5Yb(MoO_4)_4}$ recorded at 3.3 K, along with the Rietveld refinement profile obtained using TOPAS.Diffrac software. \par
$\mathrm{Na_5Yb(MoO_4)_4}$ crystallizes in the tetragonal space group \textit{I4$_1$/a} (No.\ 88), consistent with the scheelite type framework. The refined structural parameters from the NPD data are summarized in Table~\ref{TABLE}. The lattice constants of $\mathrm{Na_5Yb(MoO_4)_4}$ are $a = b$ =11.3125(1){~\AA} and $c$= 11.3316(2){~\AA}. As shown in Fig.~\ref{NPD}(a), each magnetic Yb atom is coordinated by eight oxygen atoms, forming a distorted dodecahedral environment with four Yb-O of bond length 2.322 {~\AA} and another four of 2.292 {~\AA}. The Mo atoms occupy tetrahedral sites surrounded by four oxygen atoms, while the Na atoms occupy two different crystallographic sites:  Na1 atoms (tetrahedrally co-ordinated) and Na2 atoms (octahedrally co-ordinated). \par
Each Yb atom is connected to four nearest neighboring Yb atom via two equivalent O-Mo-O bridges, giving rise to a stretched diamond lattice, as illustrated in Fig.~\ref{NPD}(b) and Fig.~(c). The nearest-neighbor distance ($J1$) is  6.33 \AA, corresponding to super-super exchange interactions mediated through two symmetric RE-O-Mo-O-RE pathways. The associated bond angles of the magnetic lattice are 101.6$^{\circ}$  and  126.8$^{\circ}$, respectively. The next-nearest ($J_{2}$) and next-next-nearest ($J_{3}$) Yb-Yb seperation are 9.80 \AA\ and 10.21 {\AA}, respectively. The extended distance between the magnetic atoms and the complex pathways effectively suppress $J_{2}$ and $J_{3}$ couplings, indicating $J_{1}$ as the only dominant exchange pathway, consistent with our  first-principle calculations.  Consequently, the system can be approximated as a three-dimensional network of very weakly coupled spins on a stretched-diamond lattice.\par
The NPD data down to 3.3 K also revealed no additional magnetic Bragg peaks, confirming the absence of long-range magnetic order in $\mathrm{Na_5Yb(MoO_4)_4}$. This suggests that the exchange interactions are too weak to drive magnetic ordering above 3 K, consistent with the magnetic and thermodynamic analyses (discussed in subsequent sections) supporting this interpretation.\par 
\begin{table}[ht]
	\caption{Refined structural parameters from the NPD data acquired at 3.3 K. 
		Space group: \textit{I4$_1$/a} (No.\ 88), $a = 11.3125(1)$~\AA, 
		$b = 11.3125(1)$~\AA, $c = 11.3316(2)$~\AA; 
		$\alpha = \beta = \gamma = 90^{\circ}$; 
		$V = 1450.12(4)$~\AA$^3$.}
	\begin{ruledtabular}
		\begin{tabular}{lcccccc}
			Atom & Site & $x$ & $y$ & $z$ & Occ & $B_\mathrm{eq}$ / \AA$^2$ \\
			\hline
			Na1 & 4b  & 0         & 0         & 0.5       & 1 & 2.5(5) \\
			Na2 & 16f & 0.1178(7)  & 0.2072(7)  & 0.7842(7)  & 1 & 2.5(5) \\
			Yb1 & 4a  & 0         & 0         & 0         & 1 & 1.3(4) \\
			Mo1 & 16f & 0.0934(4)  & 0.1835(4)  & 0.2579(3)  & 1 & 0.4(4) \\
			O1  & 16f & 0.0311(5)  & 0.8529(4)  & 0.3477(3)  & 1 & 1.2(4) \\
			O2  & 16f & 0.0611(4)  & 0.3868(5)  & 0.4097(4)  & 1 & 1.2(4) \\
			O3  & 16f & 0.0824(4)  & 0.3251(4)  & 0.1899(4)  & 1 & 1.2(4) \\
			O4  & 16f & 0.1771(3)  & 0.2853(4)  & 0.6044(6)  & 1 & 1.2(4) \\
		\end{tabular}
	\label{TABLE}
	\end{ruledtabular}
\end{table}
\subsection{Magnetization}\label{B}
The temperature dependence of  magnetic susceptibility ($\chi(T)$) for $\mathrm{Na_5Yb(MoO_4)_4}$ measured between the temperature range 1.8 K $\leq$ $T$ $\leq$ 350 K in an applied field of $\mu_{0}H$ = 0.01 T is displayed in Fig.~\ref{mag}(a). The system remains paramagnetic down to 1.8 K, with no indication of long-range magnetic ordering. Figure~\ref{mag}(b) shows the temperature dependence of the inverse magnetic susceptibility after subtracting the diamagnetic contribution, \(\chi_{\mathrm{dia}} = -2.72 \times 10^{-4}\,\mathrm{cm^3/mol}\)~\cite{Bain2008}. \\ Above \(T = 180\) K, the inverse magnetic susceptibility was fitted using the Curie--Weiss (CW) law, \(\chi(T) = C/(T-\theta_{\rm CW})\), where \(C\) is the Curie constant and \(\theta_{\rm CW}\) is the CW temperature. The solid pink line in Fig.~\ref{mag}(b) represents the corresponding CW fit, which yields \(C = 2.28~\mathrm{cm^3\,mol^{-1}\,K^{-1}}\) and \(\theta_{\rm CW}^{\rm HT} = -59.49\) K. The estimated effective magnetic moment, \(\mu_{\rm eff}^{\rm exp} = \sqrt{8C} = 4.27~\mu_{\mathrm{B}}\), is close to the theoretically expected value, \(\mu_{\rm eff}^{\rm the} = 4.53~\mu_{\mathrm{B}}\), for free \(\mathrm{Yb}^{3+}\) ions \((\ ^2F_{7/2}; S = 1/2, L = 3)\). The magnitude of the CW temperature ($|\theta_{\rm CW}^{\rm HT}| \sim 60$ K) reflects the characteristic energy scale of the crystal field excitations for the present system.  \\  It is worth noting that the experimental data  starts to deviate high-temperature CW fit (pink line in Fig.~\ref{mag}(b)) below \(\sim 60\) K, indicating the presence of a low-energy Kramers doublet state. For rare-earth Kramers ions such as \(\mathrm{Yb}^{3+}\) ions, the CEF splits the ground-state multiplet into several Kramers doublets. In the case of \(\mathrm{Yb}^{3+}\) \((4f^{13})\), the \(J = 7/2\) multiplet is typically split into four Kramers doublets, with their energies determined by the local crystal symmetry (see Sec.~\ref{CEF}). At sufficiently low temperatures, only the lowest Kramers doublet remains thermally populated, leading to an effective spin-\(\tfrac{1}{2}\) ground state.\\
To estimate the energy gap between the low-lying ground-state doublet and the first excited doublet, the inverse magnetic susceptibility data in the temperature range \(1.8\text{--}300\) K were fitted using a phenomenological two-level CEF function~\cite{BESARA201423}
\begin{equation}
\frac{1}{\chi} = 
\frac{8 (T - \theta_{\rm CW}^{\rm LT})(1 + e^{-\Delta_{10}/T})}
{\mu_{\mathrm{eff},0}^{2} + \mu_{\mathrm{eff},1}^{2} e^{-\Delta_{10}/T}}.
\label{gapcef}
\end{equation}
The fit yielded \(\mu_{\mathrm{eff},1} = 5.13~\mu_{\mathrm{B}}\) and \(\mu_{\mathrm{eff},0} = 2.99~\mu_{\mathrm{B}}\), corresponding to the effective magnetic moments of the first excited state and the ground state, respectively. The value of \(\mu_{\mathrm{eff},0} = 2.99~\mu_{\mathrm{B}}\) is significantly smaller than the high-temperature moment \(\mu_{\mathrm{eff}}^{\rm exp} = 5.3~\mu_{\mathrm{B}}\) derived from the CW fit, suggesting that the ground state is a Kramers doublet with an effective \(J_{\mathrm{eff}} = 1/2\) moment of Yb$^{3+}$ ions. The energy gap \(\Delta_{10}/k_{\rm B} = 158\) K confirms that the ground-state Kramers doublet is well isolated from the first excited doublet state. On the other hand, the negative value of \(\theta_{\rm CW}^{\rm LT} = -3.6~\mathrm{K}\) can be viewed as a possible indication of antiferromagnetic exchange interactions between the \(J_{\rm eff} = 1/2\) moments in the low-energy Kramers doublet state.\\
To further elucidate the nature of the magnetic exchange interactions, we attempted CW fitting of the low-temperature susceptibility data below 30 K over different temperature ranges. However, the fitting parameters were found to be highly sensitive to the chosen temperature window, making a reliable analysis difficult. Nevertheless, the fit performed below 8 K (Fig.~\ref{mag}(c)), where the muon spin relaxation rate becomes nearly temperature independent, gives a reasonably good description of the data with \(\mu_{\rm eff} = 2.61~\mu_{\rm B}\) and \(\theta_{\rm CW}^{\rm LT} = -0.493(19)~\mathrm{K}\). The obtained effective moment, \(\mu_{\rm eff} = 2.61~\mu_{\rm B}\), is consistent with the value typically expected for a \(J_{\rm eff} = 1/2\) Kramers doublet ground state~\cite{PhysRevB.106.104404}. The low-temperature effective moment \(\mu_{\rm eff} = 2.61~\mu_{\rm B}\) gives an effective \(g\)-factor of \(g \approx 3.01\)~\cite{PhysRevB.106.104404, PhysRevB.108.134408,PhysRevB.106.014409}.  The negative CW temperature, \(\theta_{\rm CW}^{\rm LT} = -0.493(19)~\mathrm{K}\), suggests the presence of weak antiferromagnetic exchange interactions, as commonly observed in \(4f\) rare-earth materials~\cite{PhysRevB.106.104404,PhysRevB.108.134408, kumar2024,PhysRevB.106.104408}.\\
As shown in Fig.~\ref{mag}(d), the temperature dependence of  \(ac\) magnetic susceptibility exhibits no frequency dependence, confirming the absence of spin-glass behavior, magnetic inhomogeneity, or significant anisotropy barriers down to 1.8 K. To further examine the low-temperature magnetic response, the \(ac\) magnetic susceptibility measurements were carried out from 60 mK to 4 K using drive frequencies between 400 and 4000 Hz (Fig.~\ref{mag}(e)). No frequency-dependent peak is observed over this temperature range, ruling out both long-range magnetic ordering and spin freezing down to 60 mK. A weak upturn in the susceptibility below 0.4 K is also observed, which may arise from short-range magnetic correlations.\\ 
The isothermal magnetization measured at 2 K (Fig.~\ref{mag}(f)) is well fitted by the Brillouin function,
$M(H) = M_S 
\left[
\frac{2J + 1}{2J} \coth\!\left( \frac{2J + 1}{2J} y \right)
- \frac{1}{2J} \coth\!\left( \frac{y}{2J} \right)
\right],
$
together with an additional linear term \(\chi_{\rm VV}H\), where \(y = \frac{g\mu_BJ\mu_0H}{k_{\rm B}T}\) with the Boltzmann constant $k_{\rm B}$ and \(\chi_{\rm VV}\) represents the Van Vleck susceptibility. The fit yields \(\chi_{\rm VV} = 0.031~\mu_{\rm B}/\mathrm{Yb}^{3+}/\mathrm{T}\) and a Land\'e \(g\)-factor of \(g = 2.98\), assuming \(J = 1/2\), which is consistent with the value estimated from the CW fit.
\begin{figure*}[t]
	\centering
	\includegraphics[width=0.76\textwidth]{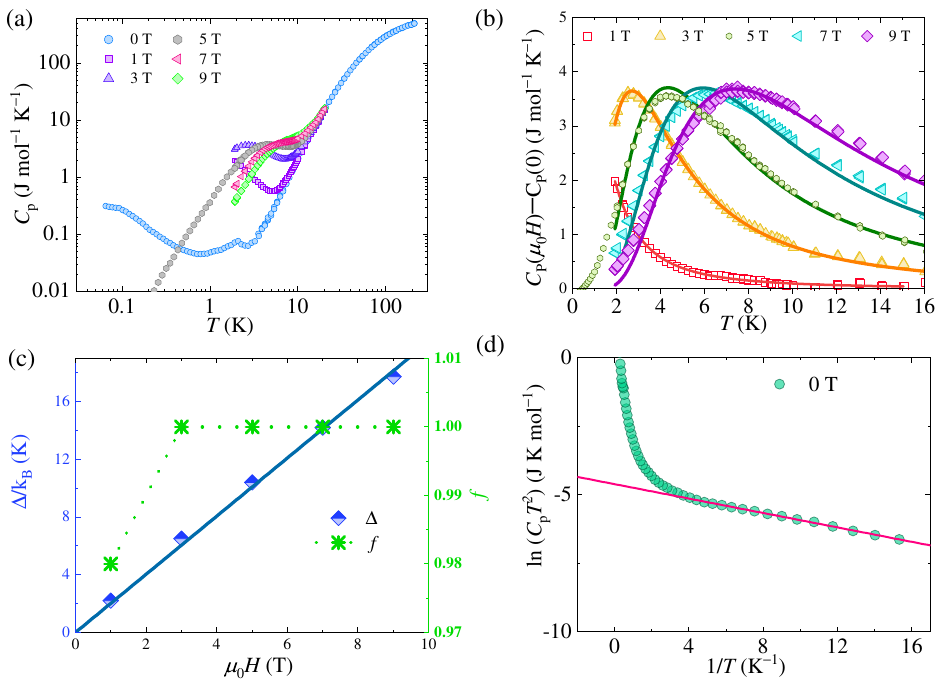}
	\caption{(a) Temperature dependence of specific heat at several magnetic fields. (b) Schottky specific heat as a function of temperature at various applied magnetic fields. (c) Magnetic-field dependence of the Zeeman gap ($\Delta/k_{\rm B}$; left y-axis) with a linear fit, along with the spin population fraction ($f$; right-y-axis). (d) Linear-fit analysis of the \(\ln(C_{\rm p}T^2)\) versus \(1/T\) plot at 0 T. Below 0.6~K, \(C_{\rm p}\) is well described by the phenomenological form \(C_{\rm p}\propto T^{-2}e^{-\Delta_{\rm dip}/T}\), giving a linear behavior in this representation and yielding \(\Delta_{\rm dip}=0.1314\)~K.  }
	\label{cp}
\end{figure*}
\subsection{Specific heat}\label{SP}
Low-temperature specific-heat (\(C_{\rm p}\)) measurements were carried out to elucidate the magnetic ground state of \ce{Na5Yb(MoO4)4}. Figure~\ref{cp}(a) shows the temperature dependence of \(C_{\rm p}\)  at several magnetic fields. In zero field, no signature of long-range magnetic order is detected down to 60 mK. Below 1 K, the \(C_{\rm p}\) data show an upturn, implying the development of short-range spin correlations, consistent with the very small CW temperature ($|\theta_{\rm CW}^{\rm LT} |= 0.49$ \text{K}). The weak kink that appears around 2.22 K is attributed to long-range magnetic ordering arising from an unavoidable minor impurity phase of \ce{Yb2O3}~\cite{PhysRevB.107.224416,PhysRevB.109.024427}.\\  At higher fields $\mu_{0}H$ $>$ 1 T, a broad maximum emerges above 2 K, shifting progressively to higher temperatures with increasing fields. This feature corresponds to the well-known Schottky anomaly, which arises from the Zeeman splitting of the lowest Kramers doublet. 
In order to quantify this effect, the Schottky contribution was evaluated as
$C_{\rm Sch}$ = $C_{\rm p}(\mu_{0}H)-C_{\rm p} (0)$
and was fitted using the two-level Schottky model (solid line in Fig.~\ref{cp}(b))
\begin{equation}
C_{\mathrm{Sch}}(T) = fR \left( \frac{\Delta}{k_{\mathrm{B}}T} \right)^{2} 
\frac{e^{\Delta / k_{\mathrm{B}}T}}{\left( 1 + e^{\Delta / k_{\mathrm{B}}T} \right)^{2}},
\label{eq:schottky2}
\end{equation}
where $R$ is the universal gas constant (8.34 J/mol.K), $f = g_{1}/g_{0}$ represents the ratio of free spins populating the higher energy level ($g_{1}$) relative to the ground state ($g_{0}$), and $\Delta/k_{\mathrm{B}}$ denotes the energy gap between the two levels.\\
The extracted fit parameters reveal that $\Delta/k_{\mathrm{B}}$ increases monotonically with the applied magnetic field (Fig.~\ref{cp}(c)). This linear increase corresponds to the Zeeman splitting between the ground state and excited state of the lowest Kramers doublet without any crossover among the excited CEF levels. Also, $\Delta/k_{\mathrm{B}} \ll 130$ K even at 9 T, indicating that $\mathrm{Na_5Yb(MoO_4)_4}$ behaves as a two-level pseudospin-$1/2$ system in higher fields. The linear fit of \(\Delta/k_{\mathrm{B}}\) as a function of magnetic field, based on the relation
$
\Delta = g \mu_{\mathrm{B}} \mu_{0} H,$
yields a Land\'e \(g\)-factor of \(g = 3\), which is close to the value obtained from the Brillouin-function fit. In addition, the free-spin factor $f$ is less than unity below 3 T. These observations hint at the presence of spin correlations at sub-kelvin temperatures. Only for magnetic fields $\geq 3$ T does $f$ approach unity, recovering  a purely Zeeman-split two-level system.\\
 As the exchange interactions are of the order of $\mu$eV, it is quite convenient  to consider that dipolar interactions may play an important role in the present case \cite{PhysRevB.106.014409,PhysRevB.104.024427}. The presence of a gapped spin-excitation spectrum arising from dipole--dipole-induced anisotropy is analyzed by fitting the low-temperature specific heat below 0.6 K with the expression
 $C_{\rm p} \propto T^{-2} e^{-\Delta_{\mathrm{dip}}/T}$, 
where \(\Delta_{\mathrm{dip}}\) represents the energy gap of the spin excitations~\cite{PhysRevLett.99.097201}. The linear fit of the $\ln(C_{\mathrm{p}}T^{2})$ vs $1/T$ plot shown in Fig.~\ref{cp}(d) provided a straight line with $\Delta_{\mathrm{dip}} = 0.1314(13)$ K. This energy gap supports the presence of gapped spin excitations in the system and may provide a possible origin of the persistent spin dynamics observed in the muon spin relaxation results discussed in the section Sec.~\ref{muSR}. 
\begin{figure*}[t]
	\centering
	\includegraphics[width=\textwidth]{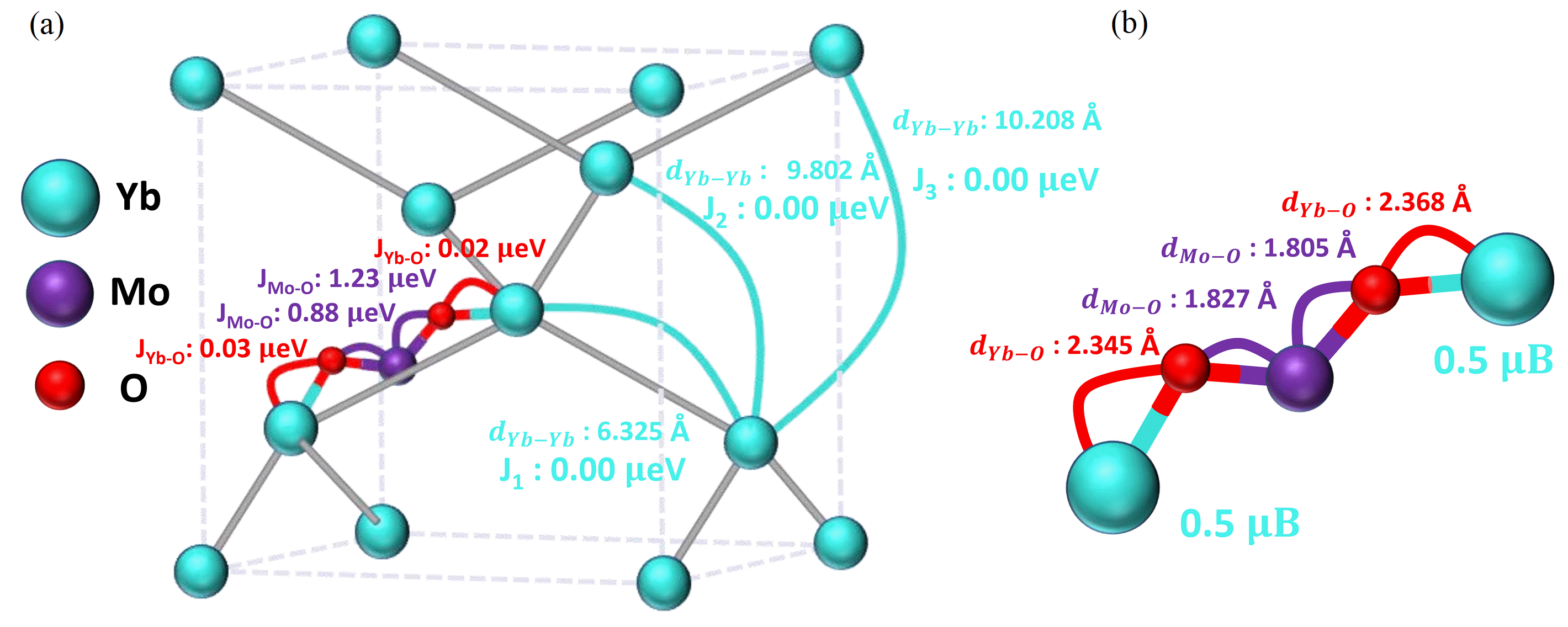}
	\caption{\label{DFT}(a) Inter-site exchange parameters ($J$) between magnetic Yb atoms as well as the exchange parameters mediated through the Yb–O–Mo–O–Yb pathway. The cyan lines indicate the interatomic distances and the exchange parameters between Yb atoms. (b) The corresponding bond lengths along the O–Mo–O pathway and the magnetic moments of Yb are also presented.}
\end{figure*} 
\begin{table*}[t]
	\caption{Lattice parameters, magnetic moments, and exchange parameters 
		calculated using PBE.}
	\label{tab:exchange}
	\begin{ruledtabular}
		\begin{tabular}{lcccccccc}
			\multicolumn{9}{c}{Na$_5$Yb(MoO$_4$)$_4$} \\[2pt]
			\hline
			Bond & Yb1--Yb2 & Yb1--Yb3 & Yb1--Yb4 & Yb--O & O--Mo & Mo--O & O--Yb \\
			& $J_{1}$ & $J_{2}$ & $J_{3}$ &  &  &  &  \\[2pt]
			\hline
			Bond length (\AA) 
			& 6.325 & 9.802 & 10.208 & 2.345 & 1.827 & 1.805 & 2.368 \\
			Exchange $J$ ($\mu$eV) 
			& 0.001 & 0.000 & 0.000 & 0.030 & 0.884 & 1.234 & 0.023 \\[4pt]
			\multicolumn{9}{l}{Magnetic moment:\quad 
				Yb: 0.495~$\mu_\mathrm{B}$ \quad 
				O1: 0.009~$\mu_\mathrm{B}$ \quad 
				O2: 0.011~$\mu_\mathrm{B}$ \quad 
				Mo: 0.002~$\mu_\mathrm{B}$ } \\[8pt]
		\end{tabular}
	\end{ruledtabular}
\end{table*}
\subsection{DFT calculations}
The experimentally observed absence of long-range magnetic order and the presence of spin fluctuations suggest weak magnetism with small inter-site exchange parameters in the compound $\mathrm{Na_5Yb(MoO_4)_4}$. To reveal the underlying mechanism, we performed first-principles electronic-structure calculations based on DFT (see Appendix \ref{APP}). The calculated magnetic moments and inter-site exchange parameters, together with the corresponding exchange paths and interatomic distances in the magnetic sublattice, are shown in Fig.~\ref{DFT}(a) and Table~\ref{tab:exchange}.

Despite the calculated nontrivial magnetic moment of approximately $0.5\,\mu_{\mathrm{B}}$ for Yb, the relatively large interatomic distance of $6.325$~\AA\ between neighboring Yb ions (Fig.~\ref{DFT}) implies negligible direct exchange interactions between magnetic ions~\cite{longf}. In $\mathrm{Na_5Yb(MoO_4)_4}$, the direct exchange parameters $J_1$, $J_2$, and $J_3$ are effectively zero. Meanwhile, the exchange parameter between Yb and neighboring oxygen ions, $J_{\mathrm{Yb\text{-}O}}$, is about $0.03\,\mu$eV (Fig.~\ref{DFT}(b) and Table~\ref{tab:exchange}). Although $J_{\mathrm{Yb\text{-}O}}$ is stronger than the direct Yb--Yb exchange parameters, it remains in the sub-$\mu$eV range, indicating extremely weak magnetism mediated through the Yb--O bonds.

On the other hand, the exchange parameter $J_{\mathrm{Mo\text{-}O}}$ is nearly $1\,\mu$eV, which is an order of magnitude larger than $J_{\mathrm{Yb\text{-}O}}$. Consequently, super-superexchange interactions could, in principle, be mediated through the O--Mo--O pathway. However, owing to the nearly vanishing value of $J_{\mathrm{Yb\text{-}O}}$, the longer-range magnetic interaction through the Yb--O--Mo--O--Yb pathway in $\mathrm{Na_5Yb(MoO_4)_4}$ is effectively suppressed at the first step (Yb--O) of the interaction chain.

This result is fully consistent with the experimentally observed absence of long-range magnetic order down to $60$~mK, as well as with the dynamic spin fluctuations observed by muon spin relaxation measurements down to $50$~mK in this Yb-based system. From a theoretical perspective, longer-range magnetic interactions at the sub-$\mu$eV energy scale correspond to an ordering temperature of approximately $0.3$~mK for $\mathrm{Na_5Yb(MoO_4)_4}$, which strongly supports our experimental observations.
 \begin{figure*}[t]
	\centering
	\includegraphics[width=0.76\textwidth]{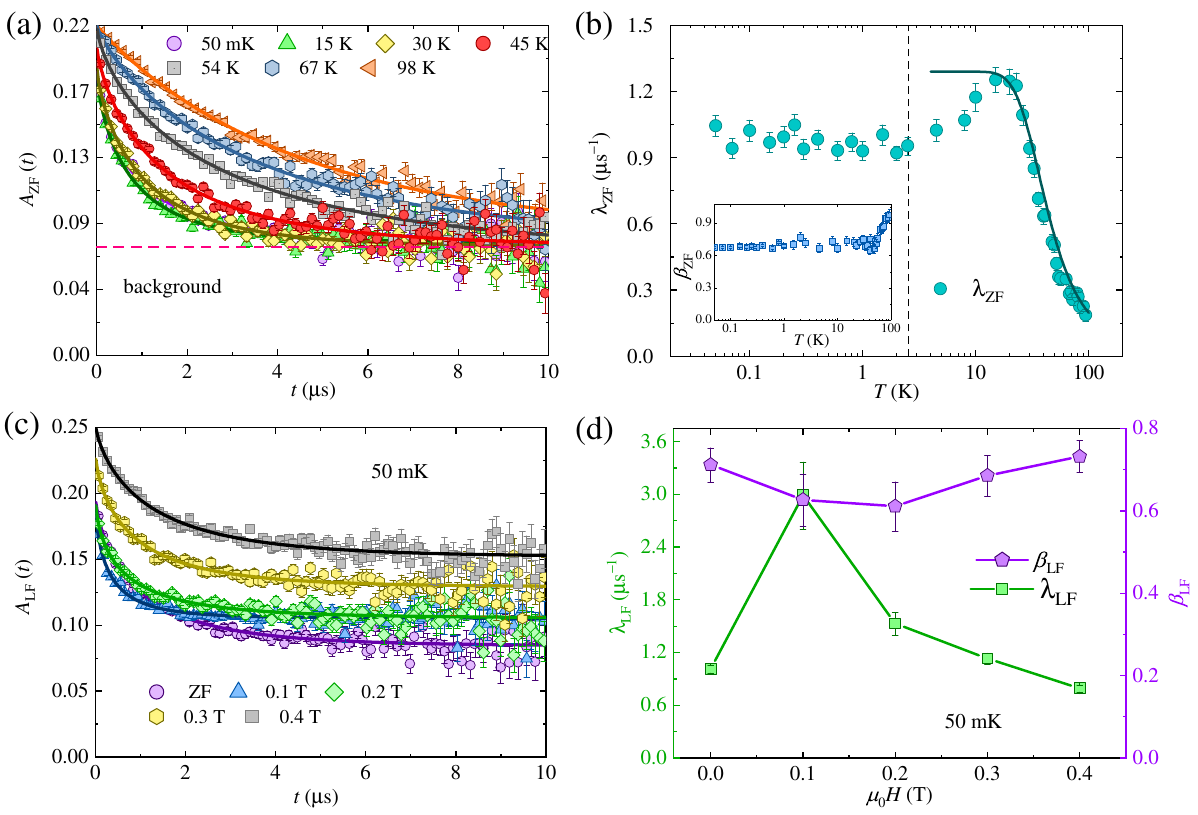}
	\caption{\label{ZF}(a) Zero-field (ZF) $\mu$SR spectra of   \ce{Na5Yb(MoO4)4} at selected temperatures. (b) Temperature dependence of the muon spin relation rate ($\lambda_{\rm ZF}$) in ZF, where the solid line represents the thermally activated behaviour of  crystal field excitations as described in the text. The inset shows the corresponding temperature dependence of the stretched exponent, $\beta_{\rm ZF}$. (c) Time evolution of the muon spin asymmetry in several longitudinal magnetic fields at $T$ = 50 mK.  (d) Longitudinal-field (LF) dependence of the muon spin relaxation rate 
		($\lambda_{\mathrm{LF}}$, left axis) and the stretched exponent  
		($\beta_{\mathrm{LF}}$, right axis) at the lowest measured temperature.
	}
\end{figure*}
 
\subsection{Muon spin relaxation}\label{muSR}
Muon spin relaxation ($\mu$SR) technique provides a highly sensitive local probe of magnetism, enabling direct access to the spatial distribution and temporal fluctuations of internal magnetic fields~\cite{le2011muon}. In this technique, when positive muons are implanted into a sample, the nearly 100 \%  spin-polarized positive muons stop at interstitial sites—typically located $\approx 1$ Å from neighboring oxygen ions in $\text{Na}_{5}\text{Yb}(\text{MoO}_{4})_{4}$. At these muon sites, the muon spins experience local magnetic fields arising from nearby electronic or nuclear moments in the sample. These fields cause the muon spins to precess and relax, leading to a time-dependent decay of the initial asymmetry, which is detected in the time evolution of $\mu$SR spectra and reflects the local magnetic environment of the sample \cite{Hillier2022}.\\
In Fig.~\ref{ZF}(a), we show the time evolution of  muon spin asymmetry  at several temperatures in zero-field (ZF). At all temperatures, the asymmetry spectra show neither well-defined spontaneous oscillations, which would indicate long-range magnetic order, nor a recovery of one-third of the initial asymmetry, as expected for a static disordered magnetic state. These observations demonstrate  the absence of static magnetic ordering  in the system down to the lowest measured temperature of 50 mK. For quantitative analysis, the ZF-$\mu$SR spectra were fitted using a single stretched exponential function,
\begin{equation}\label{EQ1}
A(t) = A_{0}\exp\!\left[-(\lambda_{\mathrm{ZF}} t)^{\beta_{\rm ZF}}\right] + A_{\rm bg},
\end{equation}
where $\lambda_{\mathrm{ZF}}$ represents the  muon spin relaxation rate in ZF, A$_{0}$ is the initial asymmetry, $A_{\rm bg}$ represents a temperature-independent background contribution, and $\beta_{\rm ZF}$ characterizes the distribution of $\lambda_{\rm ZF}$. The solid line in Fig.~\ref{ZF}(a) corresponds to the fit using Eq.~\ref{EQ1}.
\\
 Shown in Fig.~\ref{ZF}(b) is  the temperature dependence of $\lambda_{\rm ZF}$. Upon lowering the temperature below 98 K, $\lambda_{\rm ZF}$  rapidly increases  down to $14~\text{K}$. Such an enhancement of the relaxation rate at elevated temperatures is typical for various 4$f$ systems and arises from thermally activated CEF excitations. In this regime, the spin dynamics are dominated by the Orbach process, which relaxes via transitions to higher-lying CEF levels at temperatures exceeding the exchange energy scale \cite{PhysRevB.109.024427}. In many $4f$ systems, a similar temperature dependence of $\lambda_{\mathrm{ZF}}$ has been observed above the $J_{\mathrm{eff}}=\tfrac{1}{2}$ Kramers doublet ground state. 
 This behavior can be  described phenomenologically using a single-gap activation  model~\cite{le2011muon,Arh2022},
 \begin{equation}\label{or}
 \frac{1}{\lambda_{\rm ZF}} = \frac{1}{\lambda_0} + \frac{\eta}{\exp\!\left(\frac{\Delta_{\mu\mathrm{SR}}}{T}\right) - 1}.
 \end{equation}
Here, $\lambda_0$ represents the contribution from electron spin fluctuations within the Kramers doublet in the $T \rightarrow 0$ K limit, $\eta$ is the amplitude associated with the Orbach relaxation process, and $\Delta_{\mu\mathrm{SR}}$ refers to the energy separation between the ground state and the first excited CEF level.  As shown in Fig.~\ref{ZF}(b), the temperature dependence of $\lambda_{\mathrm{ZF}}$ in the range $14~\text{K} \leq T \leq 98~\text{K}$ is well described by Eq.~\ref{or}, yielding $\Delta_{\mu\mathrm{SR}}/k_{\rm B} = 110 \pm 7~\text{K}$. Noticeably, this value is comparable to $\Delta_{10} = 158$ K, as deduced from the static magnetic susceptibility. In contrast to compounds such as YbMgGaO$_4$ ~\cite{PhysRevB.106.L060401}, where a further increase of $\lambda_{\mathrm{ZF}}$ is observed due to the slowing down of spin dynamics within the Kramers doublet manifold, the present compound exhibits a distinctly different behavior. Below 14~K, $\lambda_{\mathrm{ZF}}$ decreases and deviates markedly from the single-gap Orbach model. This deviation indicates that a simple  phenomenological description is inadequate, suggesting the possible involvement of an additional low-energy spin gap or a crossover from the fast- to the slow-fluctuation regime, in which the fluctuation frequency of the local fields drops below the $\mu$SR time window. Notably, the decrease of $\lambda_{\mathrm{ZF}}$ is confined to a relatively narrow temperature window between 2.5 and 14~K. However, for $T \lesssim 2.5$~K, the temperature-independent behavior of $\lambda_{\mathrm{ZF}}$ indicates that dynamic spin fluctuations in the $J_{\rm eff}$ = 1/2 Kramers doublet state persist down to the lowest measured temperature of 50~mK~\cite{Arh2022}.\\
The inset of Fig.~\ref{ZF}(b) shows the temperature dependence of the stretched exponent $\beta_{\rm ZF}$, which characterizes the inhomogeneity in the relaxation rate distribution. At high temperatures, $\beta_{\rm ZF}$ exhibits a strong temperature dependence, decreasing from approximately 1 at 98~K to about 0.68 at 50~K. This high-$T$ decrease suggests that the distribution of $\lambda_{\rm ZF}$ is governed by thermally activated depopulation of excited CEF levels. Below 50~K, $\beta_{\rm ZF}$ becomes only weakly temperature dependent down to 2.5~K, consistent with the change in the temperature evolution of $\lambda_{\rm ZF}$. The nearly constant value of $\beta_{\rm ZF}$ below 2.5~K indicates that the system reaches the Kramers doublet ground state, while the persistent distribution of relaxation rates likely reflects competing magnetic interactions or residual low-energy spin dynamics. At very low temperatures ($T < 2.5$~K), the value of $\beta_{\rm ZF}$ ($\approx$ 0.64) remains significantly higher than the canonical spin-glass value of $\beta_{\rm ZF} = 1/3$~\cite{PhysRevLett.72.1291}, suggesting that spin freezing is unlikely in the present compound.\\
To further understand the origin of  persistent spin fluctuations below 2.5 K, $\mu$SR measurements were performed in several longitudinal fields (LFs) at 50 mK, as shown in Fig.~\ref{ZF}(c). In LF, the relaxation primarily reflects the spectral density of the fluctuations of electronic moments sensed by the implanted muons. If the nearly constant value of $\lambda_{\mathrm{ZF}} \approx 1~\mu\text{s}^{-1}$ as $T \rightarrow 0$ K originates from quasi-static spin fluctuations, the associated width of the internal field distribution would be $\lambda_{\mathrm{ZF}}(T \rightarrow 0\ \ \text{K})/\gamma_{\mu} \approx 11.7$~Oe, with $\gamma_{\mu} = 2\pi \times 135.58$~MHz/T. In this case, a LF exceeding $\sim 120$~Oe ($\approx$ 10$\times$11.7 Oe) would be sufficient to fully recover the initial asymmetry. However, in the present system, significant relaxation persists (see Fig.~\ref{ZF}(c)) even for applied LFs much larger than this value, indicating the presence of dynamic electronic spin fluctuations. \\
Akin to the ZF-$\mu$SR spectra, the LF-$\mu$SR  spectra were modeled using a stretched-exponential relaxation function together with a field-dependent background term (solid lines in Fig.~\ref{ZF}(c)). The corresponding field dependences of the  relaxation rate $\lambda_{\mathrm{LF}}$ and the stretched exponent $\beta_{\rm LF}$ are shown in Fig.~\ref{ZF}(d).  Notably, at an applied field of $\mu_{0}H_{\mathrm{LF}} = 0.1$~T, $\lambda_{\mathrm{LF}}$ is approximately three times larger than the ZF relaxation rate at $T$ = 50 mK. This enhancement suggests the presence of weak exchange interactions, whereby the applied magnetic field lifts the degeneracy of the Kramers doublet and opens a Zeeman gap that shifts the spectral density of the spin fluctuations into the $\mu$SR frequency window. This is also reflected in the field dependence of $\beta_{\rm LF}$. Upon further increasing the applied LF, the relaxation rate gradually decreases, consistent with a progressive suppression of low-energy spin fluctuations as the Zeeman gap increases, resulting in a reduction in the density of states available for muon spin-flip processes at the Larmor frequency.\\
\begin{figure*}[t]
	\centering
	\includegraphics[width=\textwidth]{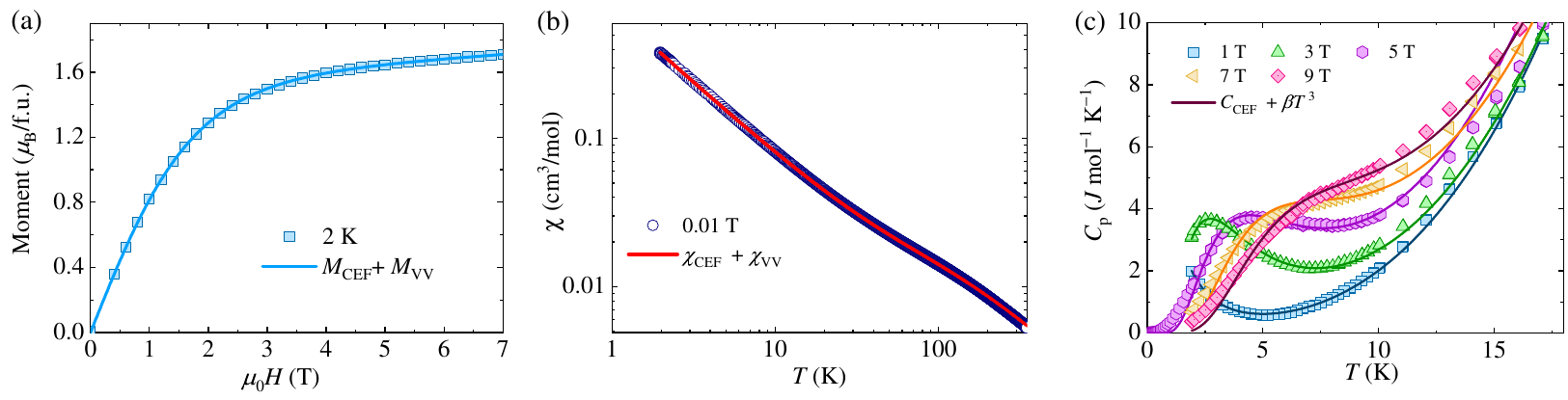}
	\caption{Field dependence of the isothermal magnetization at 2 K with the CEF-model fit.
		(b) Temperature dependence of the magnetic susceptibility along with the simulated magnetic susceptibility obtained from the crystal electric field model including a constant term.
		(c) Temperature dependence of the specific heat, fitted using the CEF model plus a $\beta T^{3}$
		term. 	}\label{PCM}
\end{figure*}
\subsection{Crystal electric field analysis}\label{CEF}
To further understand the CEF doublets associated with single-ion effects, we employed the point-charge model~\cite{HUTCHINGS1964227}. According to Hund's rules, the ground-state multiplet of the \(\mathrm{Yb}^{3+}\) ion is \({}^{2}F_{7/2}\) with total angular momentum \(J = 7/2\). In the presence of the crystalline electric field, this multiplet splits into \(2J+1 = 8\) states, which, for a Kramers ion such as \(\mathrm{Yb}^{3+}\), form four Kramers doublets.

The crystal-field Hamiltonian is expressed as
\begin{equation}
\mathcal{H}_{\mathrm{CEF}} = \sum_{l,m} B_l^m \, O_l^m(J),
\end{equation}
where \(B_l^m\) are the crystal-field parameters and \(O_l^m(J)\) are the Stevens operator equivalents expressed in terms of the total angular momentum operators~\cite{KWHStevens1952}. Here, \(l\) denotes the rank (degree) of the operator, taking values \(l = 2, 4, 6\) for \(4f\) electrons, while \(m\) labels the component and runs from \(-l\) to \(+l\). The set of parameters \(B_l^m\) encapsulates the strength and symmetry of the local crystalline environment surrounding the magnetic ion. The nonvanishing crystal-field parameters \(B_l^m\) are determined by the local site symmetry of the magnetic ion.\\
Using the point-charge model, the nonzero crystal-field parameters of the crystal-field Hamiltonian were obtained and are listed in Table~\ref{tab:CEF}. These results indicate that the local environment of the \(\mathrm{Yb}^{3+}\) ion is close to \(S_{4}\) site symmetry in Na$_{5}$Yb(MoO$_{4}$)$_{4}$. The diagonalization of the crystal-field Hamiltonian reveals four Kramers-doublet levels located at 0, 230.8, 519.5, and 555.3 K. The large separation between the ground-state doublet and the first excited doublet indicates a well-isolated Kramers-doublet ground state in Na$_{5}$Yb(MoO$_{4}$)$_{4}$, consistent with the thermodynamic results (see Sec.~\ref{B}).\\
Next, in order to analyze the thermodynamic data above 2 K, which is well above the low-temperature CW temperature and hence away from the correlated regime, we consider the single-ion Hamiltonian
\begin{equation}
\mathcal{H} = \mathcal{H}_{\rm CEF} - g_{\rm J}\mu_{\rm B}\,\mathbf{J}\cdot\mathbf{H}.
\end{equation}
The corresponding magnetization along the \(\alpha\) ($x$, $y$, and $z$) direction is given by
$
M_{\alpha} = g_{\rm J}\mu_{\rm B}\,\langle J_{\alpha} \rangle,
$
where the thermal expectation value is
$\langle J_{\alpha} \rangle
=
\frac{1}{Z}
\sum_{i}
\langle i | J_{\alpha} | i \rangle
\,e^{-E_i/k_{\rm B}T},$
with the partition function
$
Z = \sum_i e^{-E_i/k_{\rm B}T}.$ Here, \(E_i\) and \(|i\rangle\) are the eigenvalues and eigenstates of the full Hamiltonian \(\mathcal{H}\), respectively.\\ The magnetization was calculated using the crystal-electric-field Hamiltonian $\mathcal{H}$,
where the powder-averaged CEF magnetization was obtained numerically using the pycrystalfield
python package~\cite{Scheie:in5044,NewmanNg2000}. The experimental data were fitted by
$M(H)=M_{\rm CEF}(H)+ \chi_{\rm VV}H$,
where \(\chi_{\rm VV}\) = 0.019 $\mu_{\rm B}/ \text{T}$ is the temperature-independent Van Vleck contribution. 
\begin{table}[t]
	\caption{Crystal-electric-field parameters \(B_l^{m}\) for \(\mathrm{Yb}^{3+}\) ions in \ce{Na5Yb(MoO4)4}.}
	\label{tab:CEF}
	\centering
\begin{tabular}{cc}
	\hline\hline
	Parameter & Value (meV) \\
	\hline
	\(B_2^{0}\)   & \(3.510 \times 10^{-1}\) \\
	\(B_4^{-4}\)  & \(-1.997 \times 10^{-1}\) \\
	\(B_4^{0}\)   & \(9.245 \times 10^{-3}\) \\
	\(B_4^{4}\)   & \(4.517 \times 10^{-2}\) \\
	\(B_6^{-4}\)  & \(1.421 \times 10^{-3}\) \\
	\(B_6^{0}\)   & \(-1.011 \times 10^{-4}\) \\
	\(B_6^{4}\)   & \(-6.269 \times 10^{-4}\) \\
	\hline\hline
\end{tabular}
\end{table}
 The good agreement between the experimental data and the fitted curve confirms that the magnetization is well described by the single-ion CEF model with an additional linear contribution.\\
 To account for the typical response of the polycrystalline sample, the magnetic susceptibility was averaged over the three principal crystallographic directions, i.e.,
$
 \chi_{\rm powder}(T) = \frac{1}{3}\left(\chi_x + \chi_y + \chi_z\right),
$
 where \(\chi_{\alpha} = M_{\alpha}/H\).
 The experimental magnetic susceptibility at 0.01 T was fitted using the relation
$ \chi(T) = \chi_{\rm CEF}(T) + \chi_{\rm VV},$
 where \(\chi_{\rm VV}\) = 3.64$\times$ 10$^{-4}$ cm$^{3}$/mol represents the temperature-independent Van Vleck contribution. The corresponding fitted curve is presented in Fig.~\ref{PCM}(b).
\\
As discussed in Sec.~\ref{SP}, the broad maximum in \(C_{\rm p}\) at finite magnetic fields originates from the Zeeman splitting of the \(J_{\rm eff}=1/2\) Kramers-doublet ground state. Accordingly, the low-temperature specific heat \(C_{\rm p}(T,H)\) was analyzed within an effective \(J_{\rm eff}=1/2\) framework, with the Land\'e factor taken as an effective parameter \(g_{\rm eff}\) and treated as a fitting parameter. The specific heat was calculated from the field-dependent eigenvalues \(E_n(H)\) as
\begin{equation}
\begin{aligned}
C_{\rm CEF}(T,H) &= \frac{R}{(k_{\rm B}T)^2}
\Biggl[
\frac{\sum_n E_n^2 e^{-E_n/k_{\rm B}T}}{\sum_n e^{-E_n/k_{\rm B}T}}
\\
&\qquad\qquad
-
\left(
\frac{\sum_n E_n e^{-E_n/k_{\rm B}T}}{\sum_n e^{-E_n/k_{\rm B}T}}
\right)^2
\Biggr].
\end{aligned}
\end{equation}

Next, the $C_{\rm p}$ data were fitted using
$C_{\rm p}(T) =\,C_{\rm CEF}(T,H) + \beta T^3$,
where  \(\beta T^3\) represents the phonon contribution. As shown in Fig.~\ref{PCM}(c), the experimental \(C_p\) data are well reproduced by the fitted model, which yields \(g_{\rm eff} = 3.12\), consistent with the thermodynamic results.
\section{Discussion}
Na$_5$Yb(MoO$_4$)$_4$ possesses a unique stretched diamond magnetic lattice, characterized by an unusually large interatomic separation of 6.33~\AA{} between nearest-neighbor and 9.8~\AA{} between next-nearest-neighbor magnetic Yb ions.
The magnetization results reveal a Kramers-doublet ground state of the Yb$^{3+}$ ions, characterized by low-energy \(J_{\rm eff}=1/2\) moments, which is well separated from the first excited state. We note that the discrepancy in the energy gap to the first excited state obtained from magnetization, \(\mu\)SR, and point-charge model calculations likely arises from the inherent limitations of each approach.\\
 DFT calculations demonstrate that the direct exchange coupling between Yb atoms is negligible due to these large interatomic distances. Although exchange interactions may, in principle, be mediated via the extended O--Mo--O super-superexchange pathway, the calculated exchange parameters are estimated to be in the sub-$\mu$eV range, which would suppress any exchange-driven long-range magnetic order to temperatures well below 1~mK. While stretched diamond lattices can, in general, host magnetic frustration through competing $J_1$--$J_2$ exchange interactions, as reported for other Yb-based compounds~\cite{PhysRevB.103.024430,PhysRevLett.130.166703,PhysRevMaterials.6.044410}, the unusually large Yb--Yb separations in the present system render both $J_1$ and $J_2$ effectively negligible. As a result, Na$_5$Yb(MoO$_4$)$_4$ realizes an essentially unfrustrated bipartite magnetic lattice.\par

Conventional bipartite diamond lattices are typically expected to undergo N\'eel-type magnetic ordering~\cite{oitmaa2019frustrated,PhysRevB.96.064413}. However, magnetic susceptibility, specific heat, and $\mu$SR measurements for Na$_5$Yb(MoO$_4$)$_4$ reveal no evidence of long-range magnetic order down to 50~mK. In the absence of appreciable exchange interactions, long-range dipole--dipole interactions therefore constitute the dominant inter-site coupling and are expected to govern the low-temperature magnetic behavior. Quantum dipolar magnets dominated by dipolar interactions have been identified in several rare-earth-based systems, including Ba$_3$Yb(BO$_3$)$_3$ \cite{PhysRevB.106.014409}, Gd$_2$Sn$_2$O$_7$\cite{PhysRevLett.99.097201}, and Yb$_3$Ga$_5$O$_{12}$\cite{PhysRevB.104.024427}.

The characteristic nearest-neighbor dipolar interaction energy $D_{\mathrm{nn}}$ can be estimated as~\cite{https://doi.org/10.1002/qute.201900089}
\begin{equation}
D_{\mathrm{nn}}/k_{\rm B} \approx \frac{\mu_0}{4\pi k_{\rm B}}\frac{\mu_{\mathrm{eff}}^2}{r_{\mathrm{nn}}^3},
\end{equation}
where $\mu_{\mathrm{eff}}$ is the effective magnetic moment and $r_{\mathrm{nn}}$ is the nearest-neighbor distance. Substituting $r_{\mathrm{nn}} = 6.33$~\AA{} and $\mu_{\mathrm{eff}}$ = 1.58~$\mu_{\rm B}$, we obtain $D_{\mathrm{nn}} \approx 0.006$~K. Despite this small  energy scale, the specific heat exhibits a pronounced upturn below approximately 1 K. This apparent discrepancy reflects the fact that $D_{\mathrm{nn}}$ represents a local interaction scale, whereas the low-temperature specific heat is determined by the root-mean-square distribution of internal dipolar fields arising from the long-range nature of dipole--dipole interactions. Contributions from multiple coordination shells enhance the effective dipolar energy scale, allowing thermodynamic signatures of dipolar correlations to emerge at temperatures substantially higher than $D_{\mathrm{nn}}$. 

Consistently, the low-temperature specific heat is well described by a gapped form with an extracted gap $\Delta_{\mathrm{dip}} \approx 0.1314$~K. This gap represents an emergent collective dipolar energy scale associated with the root-mean-square internal field generated by long-range dipolar correlations, further enhanced by strong single-ion anisotropy and weak exchange coupling. The onset of the specific heat upturn below 1 K is therefore governed by this collective dipolar scale rather than by the nearest-neighbor dipolar interaction alone. We note that these gapped excitations reflect conventional dipolar correlations of the weakly interacting Yb$^{3+}$ moments and should not be interpreted as exotic excitations as seen in frustrated quantum magnets or spin liquids.\par

Further, $\mu$SR measurements reveal persistent spin dynamics with no evidence of static internal magnetic fields between 50~mK and 2.5~K, demonstrating the absence of spin freezing or long-range magnetic order down to the lowest measured temperatures.  These observations indicate that dipolar interactions do not induce long-range magnetic order even on a bipartite lattice, and that the Yb moments are not isolated spins governed solely by single-ion anisotropy. Rather, the system realizes a weakly correlated dipolar paramagnetic state in which long-range dipolar interactions dominate the low-temperature magnetic ground state but remain insufficient to stabilize static order in the absence of appreciable exchange coupling.\par

From an applied perspective, magnetic systems of this type are of interest for potential adiabatic demagnetization refrigeration (ADR) applications~\cite{Tokiwa2021,app16010290}. The key requirements for paramagnetic ADR materials include extremely low magnetic ordering temperatures, high magnetic ion density and entropy per unit volume, a large field-tunable magnetic entropy change, minimal lattice and nuclear specific heat contributions (particularly for mK-ADR), as well as good thermal conductivity and chemical and mechanical stability~\cite{WIKUS2014150}. Conventionally, hydrated paramagnetic salts such as cerium magnesium nitrate (CMN), ferric ammonium alum (FAA), and chromium potash alum (CPA) have been widely employed as ADR refrigerants. In these systems, weak exchange interactions arise from the large separation between magnetic ions mediated by crystallization water~\cite{WIKUS2014150}. The magnetic ordering temperatures of CMN, FAA, and CPA are approximately 2, 30, and 10~mK, respectively, with corresponding volumetric magnetic entropy densities of 16, 53, and 42~mJ\,K$^{-1}$\,cm$^{-3}$~\cite{PhysRev.148.509,fisher1973,daniels1954}. Despite their historical importance, these materials suffer from limited chemical stability due to the presence of temperature-sensitive crystal water molecules~\cite{WIKUS2014150,daniels1954}.\\ \\
More recently, rare-earth-based quantum disordered magnets, including frustrated and spin-liquid candidates, have been explored as potential alternatives to conventional hydrated salts~\cite{arjun2023adiabatic,Tokiwa2021,AJesche2023,XYLiu2022}. In this context, Na$_5$Yb(MoO$_4$)$_4$ may be regarded as a promising candidate for further exploration as an ADR material. The large separation between Yb ions and the absence of crystallization water confer enhanced chemical stability during repeated ADR cycles, while the absence of magnetic order down to 50~mK, a magnetic ion density of approximately 2.8~nm$^{-3}$, and a volumetric magnetic entropy density of about 26.4~mJ\,K$^{-1}$\,cm$^{-3}$ are comparable to those of conventional hydrated paramagnets. \section{Conclusion}
The crystal structure and magnetic ground state of polycrystalline Na$_5$Yb(MoO$_4$)$_4$ were comprehensively investigated using powder diffraction, magnetization, specific heat, muon spin relaxation  measurements, and PBE+$U$ calculations. These investigations consistently indicate that Na$_5$Yb(MoO$_4$)$_4$ exhibits a very weak, dynamically correlated dipolar quantum paramagnetic state. The absence of magnetic ordering down to the lowest measured temperatures can be understood as a consequence of the combined effects of strong single-ion anisotropy of the $J_{\rm eff}=1/2$ Yb moments, dominant dipolar interactions, and extremely weak exchange coupling. While dipolar interactions set the leading energy scale among Yb-Yb spins, their long-range and anisotropic nature, together with the constrained spin degrees of freedom imposed by single-ion anisotropy, does not favor the stabilization of long-range magnetic order. The negligible exchange coupling further prevents the system from achieving an ordered ground state. As a result, the Yb moments remain dynamically correlated, forming a weakly correlated dipolar paramagnetic state rather than a magnetically ordered phase.
    \section*{Acknowledgments}
    This work is supported by the National Science and Technology Council in Taiwan (Grants No. 111–2112-M-002–044-MY3 and No. 112–2124-M-002–012), the Featured Research Center Program within the framework of the Higher Education Sprout Project by the Ministry of Education in Taiwan (Grants No. 113L9008), and Academia Sinica with Project No. AS-iMATE-113-12. R.S acknowledges the financial support provided by the Ministry of Science and Technology in Taiwan under project numbers NSTC-114-2124-M-001-009 and NSTC-113-2112-M-001-045-MY3, and Academia Sinica for the budget of AS-iMATE-115-14.The work in SKKU was supported by the National Research Foundation (NRF) of Korea (grant number RS2023-00209121, 2020R1A5A1016518). A.M.S thanks the FRC/URC of UJ, and the SA-NRF (105702) for financial assistance.  
     \section{DATA AVAILABILITY}
    The data supporting the findings of this study are available from the corresponding author upon reasonable request.
    \appendix
     \section{DFT calculations}\label{APP}
\begin{figure*}
     	\centering
     	\includegraphics[width=\textwidth]{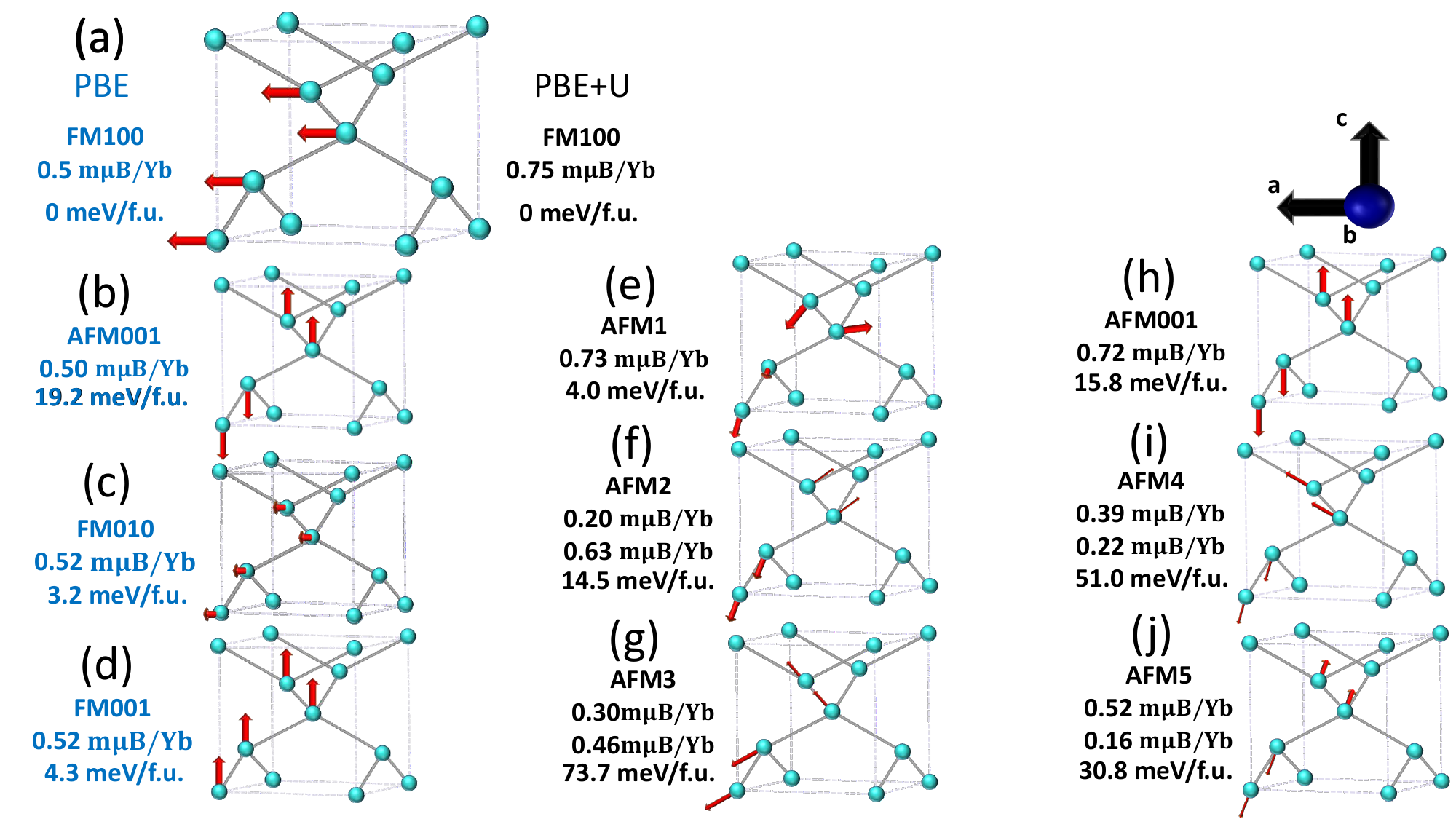}
     	\caption{Comparison of total energy and magnetic moment per magnetic atom for different magnetic orders in Na\textsubscript{5}Yb(MoO\textsubscript{4})\textsubscript{4}. \textbf{(a)} The ferromagnetic (FM) state along the (100) direction, identified as the ground state in both PBE and PBE+U calculations. \textbf{(b–d)} Alternative magnetic configurations with higher total energies in the PBE calculations, where the antiferromagnetic (AFM) state along the (001) direction exhibits the lowest energy among the AFM states. \textbf{(e–j)} Alternative magnetic configurations with higher total energies in the PBE+U calculations, with AFM1 identified as the lowest-energy AFM state. All calculations include spin-orbit coupling (SOC). }
     	\label{APP_s1}
     \end{figure*}

     The calculated magnetic configurations reveal nearly degenerate
     ground states under SOC. Both PBE and PBE+U predict FM100 as the ground
     state, with AFM001 (PBE) and AFM1 (PBE+U) being the lowest-energy AFM states.
     The arrows in Fig.~\ref{APP_s1} denote the orientation of magnetization.\par
     \begin{figure*}
     	\centering
     	\includegraphics[width=0.85\textwidth]{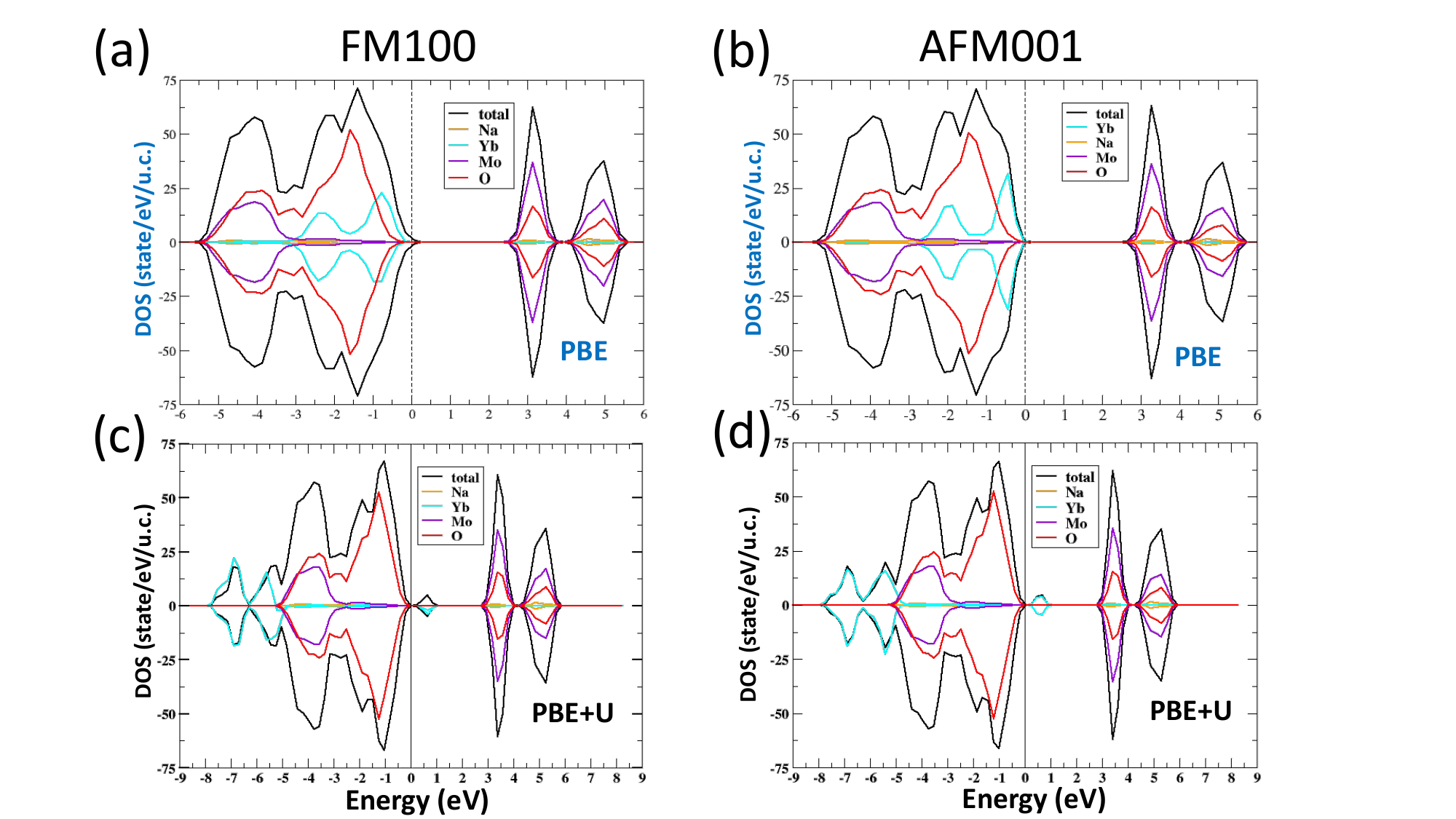}
     	\caption{Density of states (DOS) of \textbf{(a)} DOS of the FM100 state obtained from the PBE calculation. \textbf{(b)}
     		DOS of the AFM001 state obtained from the PBE calculation. \textbf{(c)} DOS of the FM100 state obtained
     		from the PBE+U calculation. \textbf{(d)} DOS of the AFM001 state obtained from the PBE+U calculation. All
     		DOS calculations are with SOC.}
     	\label{APP_s2}
     \end{figure*}
     It can be seen that the inclusion of the Hubbard U correction reduces the energy difference between ferromagnetic and antiferromagnetic. That is, the absence of other FM order in the PBE+U calculation may also stem from the inclusion of Hubbard U correction. It is worth noting that PBE+U calculations are not able to obtain converged FM010 and FM001 magnetic phases, to some extent consistent with the experimental observation of spin fluctuation in the muon spin relaxation.\par
     
     The density of states (DOS) (Fig.~\ref{APP_s2}) calculations further indicate that the inclusion of the Hubbard U term substantially modifies the dispersion of the f-orbitals for Yb atoms, particularly in the FM100 and AFM001 configurations, which is the ground state and the AFM state with lowest energy  respectively.
     \clearpage   
\bibliography{References}
\end{document}